\documentstyle[aps,prl]{revtex}

\begin{document}
\title{Carrier relaxation dynamics in intra-gap states: the case of the
superconductor YBa$_{2}$Cu$_{3}$O$_{7-\delta }$ and the charge-density-wave
semiconductor K$_{0.3}$MoO$_{3}$.}
\author{V.V.Kabanov, J.Demsar and D. Mihailovic}
\address{Solid State Physics Department, ''Jozef Stefan'' Institute, Jamova 39, 1001\\
Ljubljana, Slovenia}
\date{\today }
\maketitle

\begin{abstract}
The unusual slow carrier relaxation dynamics - observed in femtosecond
pump-probe experiments on high-temperature superconductors and recently also
in a charge-density-wave system - is analyzed in terms of a model for
relaxation of carriers in intra-gap states. The data on YBa$_{2}$Cu$_{3}$O$%
_{7-\delta }$ near optimum doping and K$_{0.3}$MoO$_{3}$ are found to be
described very well with the model using a BCS-like gap which closes at $%
T_{c}$. From the analysis of the data we conclude that a significant
intra-gap density of localized states exists in these materials, which can
be clearly distinguished from quasiparticle states by the time-resolved
optical experiments because of the different time- and
temperature-dependences of the photoinduced transmission or reflection.
Localized charges are suggested to be the most likely origin of the
intra-gap states, while the similarity of the response in the two materials
appear to exclude spin and vortex excitations.
\end{abstract}

\twocolumn

\section{Introduction}

Photoinduced absorption or reflection spectroscopy using femtosecond lasers
is potentially a very powerful tool for the study of the electronic
structure of superconductors and related materials. Recent pump-probe
experiments on cuprate superconductors \cite{Han,Stevens,Demsar,kdbm} and
some other materials \cite{bb} have shown that a photoinduced change in
absorption or reflection can be observed at low temperatures and especially
for $T<T_{c}$. The effects are believed to be caused by excited state
absorption of the probe pulse from photoexcited quasiparticle (QP) states 
\cite{kdbm} and theoretical analysis of the response was found to be in good
agreement with experimental data on YBa$_{2}$Cu$_{3}$O$_{7-\delta }$ over a
wide range of doping \cite{kdbm}. However, in addition to the QP response
which occurs on the picosecond and subpicosecond timescale, a distinct
slower response was also consistently observed in YBa$_{2}$Cu$_{3}$O$%
_{7-\delta }$ (YBCO) \cite{Stevens}, Bi$_{2}$Sr$_{2}$CaCu$_{2}$O$_{8}$
(BISCO) and Bi$_{2}$Y$_{x}$Ca$_{1-x}$SrCu$_{2}$O$_{8}$ \cite{Thomas} and
more recently in the charge-density-wave (CDW) quasi one-dimensional Peierls
insulator K$_{0.3}$MoO$_{3}$\cite{bb}. It was thought to be of non-thermal
origin (detailed analysis is given in Ref. \cite{Thomas}) and occurs on a
timescale of 10$^{-8}$ s or longer (see Fig. 1 for details). Its anomalous $%
T $-dependence \cite{Stevens} - which is qualitatively different from the $T$%
-dependence of the fast QP recombination dynamics - lead the authors to the
suggestion that the signal is due to localized states near the Fermi energy.
However, the processes involved were so far not discussed in any detail.

In this paper we examine quantitatively the photoinduced absorption (or
reflection) from localized intra-gap states. We develop a theoretical model
for two different cases: (i) the case of a material with a BCS-like
collective gap and (ii) the case of a $T$-independent gap (sometimes called
a pseudogap) which exists above $T_{c}$. We compare the model predictions
with\ the available data on the temperature-dependence of the photoinduced
relaxation on the nanosecond timescale in two different materials, both of
which are generally thought to have a low-temperature gap: YBa$_{2}$Cu$_{3}$O%
$_{6.9}$ ($T_{c}$ = 90K) and K$_{0.3}$MoO$_{3}$ ($T_{c}=183$K).

\section{Theoretical model}

As the name implies, pump-probe spectroscopy involves the excitation of the
material by an ultrashort pump laser pulse and the subsequent measurement of
the resulting {\em change} in optical absorption, transmission or reflection
of the sample caused by photoexcited charge carriers. As photons from the
pump laser pulse are absorbed, they excite electrons and holes in the
material (see schematic diagram in Fig.2 a)). These particles release their
extra kinetic energy by scattering amongst themselves and with phonons (step
2 in Fig. 2a)). This energy relaxation process is very rapid and the
particles end up in QP states near the Fermi energy ($E_{F}$) within $\tau
_{e}=$10$\sim $100 fs \cite{kdbm}. Subsequent relaxation is slowed down by
the presence of the gap and a relaxation bottleneck is formed. From
pump-probe photoinduced transmission experiments in YBCO \cite{Han,kdbm} and
BISCO \cite{Thomas}, the relaxation times of the photoexcited QPs were found
to be in the range $\tau _{QP}=$0.3$\sim $3 ps.

In addition to the picosecond transient, the signal on nanosecond timescale
has been consistently observed in HTSC \cite{Stevens,Thomas,unpubdata} and
recently also in quasi 1D CDW insulator K$_{0.3}$MoO$_{3}$ \cite{bb}. In
Fig. 1 the photoinduced transient taken on YBCO at T=80 K is shown
(squares). After photoexcitation ($t=0$ ) the signal relaxes within 10 ps to
some non-zero value, that can be represented by the constant on the 100 ps
timescale \cite{Thomas}. The lifetime of the slow component, $\tau _{L},$
cannot be directly measured, since it appears to be longer than the
inter-pulse separation of $t_{r}\simeq 10-12$ ns. This results in the signal
pile-up due to accumulation of the response from many pulses. The magnitude
of this pile-up, $A$, given by the difference in the signal amplitude when
the pump pulse is unblocked (2), and zero signal, when the pump is blocked
(1) at negative time delays (see Fig. 1), is several times larger than
single pulse contribution, $a_{0},$ given be the difference between (4) and
(2).

Assuming - for simplicity - that relaxation of the signal is exponential, we
can write an equation for the steady state amplitude $A$: 
\begin{equation}
A=a_{0}\sum_{n=0}^{\infty }exp(-nt_{r}/\tau _{L})=a_{0}/(1-exp(-t_{r}/\tau
_{L}))
\end{equation}
Here $a_{0}$ is the amplitude of the signal from a single pulse and $\tau
_{L}$ is the relaxation time. Usually, $A\gg a_{0}$ and we can expand the
exponent in the denominator of Eq.(1), to obtain $\tau _{L}=t_{r}(A/a_{0})$.
From the experiments \cite{Stevens,bb} it appears that $\tau _{L}>10^{-7}$
s. This is long in comparison with the phonon relaxation time and with the
phonon escape time from the excitation volume into the bulk or thin film
substrate, which is typically 10$^{-10}$s, so we can ignore phonon escape
effects and discuss only intrinsic relaxation processes.

The process giving rise to the actual\ photoinduced optical signal from
intra-gap states is shown as step 3 in Fig.2a). Here we do not discuss the
optical probe process in detail, but make the very general assumption that
the photoinduced {\it change }in sample transmission $\Delta {\it T}/{\it T}$
or reflection $\Delta {\cal R}/{\cal R}$ is proportional to the photoinduced
density of localized states populated by the laser pump pulse. The
photoinduced signal $\Delta {\it T}/{\it T}$ (or $\Delta {\cal R}/{\cal R)}$
\ is then proportional to the number of filled localized states and the $T$%
-dependence is mainly determined by the occupation of the intra-gap states.

Since the relaxation time $\tau _{L}$ is long in comparison with $\tau _{QP}$
and phonon relaxation times, we assume that phonons and quasiparticles can
be described by equilibrium densities $N_{\omega }$ and $N$ respectively.
For the relaxation of the localized carriers we apply arguments similar to
those originally proposed by Rothwarf and Taylor \cite{Roth}. The rate
equation for the total density of localized excitations $N_{L}$ is then
given by: 
\begin{equation}
\frac{dN_{L}}{dt}=-RN_{L}^{2}-\tilde{\gamma}N_{L}+\gamma N+\beta N_{\omega }.
\end{equation}
The first term in Eq.(2) describes the recombination of two localized
excitations to a Cooper pair with a recombination rate $R$. The second and
the third terms describe the exchange of an electron or a hole between the
localized and quasiparticle states with density $N_{L}$ and $N$ respectively
and with a rates $\tilde{\gamma}/\gamma \propto \exp {(-\Delta E/k}_{B}{T)}$
where $\Delta E$ is the energy barrier between trapped carriers and
quasiparticles \cite{Ryvkin}. The last term describes the spontaneous
creation of localized excitations by phonons with a relaxation rate $\beta $%
. The four processes in Eq. (2) are shown schematically in Fig.2b)).

Assuming the ansatz for $N_{L}=N_{L0}+n_{L},$ where $N_{L0}$ is equilibrium
density of localized particles and $n_{L}$ is the photoinduced density
created by the laser pulse and taking into account that $N_{\omega }$ and $N$
are given by their equilibrium values, we can rewrite Eq.(2):

\begin{equation}
\frac{dn_{L}}{dt}=-Rn_{L}^{2}-(2RN_{L0}+\tilde{\gamma})n_{L}
\end{equation}
This equation is sufficiently general that it can be applied to different
superconductors and different gaps. (For the case of a CDW gap, the Cooper
pairs are replaced by $e-h$ pairs.) In this paper we consider a) a
superconductor with a BCS-like gap (which can also be used in the case of a
gap formed by a CDW) and b) a Bose condensate of pre-formed pairs with a $T$%
-independent gap. In the latter case, the gap is better considered as an
energy-level splitting between paired and unpaired states, and the main
difference is that the gap is $T$-independent and exists above $T_{c}$. A
general exact solution to Eq.(3) is given in the Appendix. Next we consider
the limiting behavior at low $T$ and near $T_{c}$.

In the case of a collective BCS-like gap, $\Delta _{s}\left( T\right) $ ,
the recombination rate below $T_{c}$ is to the lowest order in $\Delta _{s}$
proportional to the square of the order parameter \cite{Ovch,Kapl}: 
\begin{equation}
R\simeq \alpha (\Delta _{s}(T)/\Omega _{c})^{2}
\end{equation}
where $\alpha $ is a constant, and $\Omega _{c}$ is the phonon spectrum
cutoff frequency. (Above $T_{c}$ all recombination processes disappear as $%
\Delta _{s}(T)\rightarrow 0.)$ To relax the carriers in localized intra-gap
states via quasiparticle states (the term $\tilde{\gamma}n_{L}$ in Eq.(3)),
an energy of the order of $\Delta _{s}$ is required and so this process is
exponentially suppressed at low temperatures. The term proportional to $%
N_{L0}$ is also small at low temperatures, because the number of thermally
excited localized excitations is small as $(k_{B}T/\Delta _{s}(0))^{\mu }$,
where $\mu $ depends on the density of localized states. Therefore, for
analysis of the relaxation of localized excitations at low temperatures $%
T\ll T_{c}$, we retain only the first term in Eq.(3) giving a solution of
the form:

\begin{equation}
n_{L}(t)=n_{L}(0)/(n_{L}(0)Rt+1)
\end{equation}

To obtain the stationary solution for a repetitive laser pump pulse train
excitation, we use the condition that the total number of localized
excitations that recombine between two laser pulses should equal the number
of localized excitations created by each laser pulse:

\begin{equation}
n_{L}(0)-n_{L}(0)/(n_{L}(0)Rt_{r}+1)=\eta n_{ph}(T)
\end{equation}
where $\eta \propto \gamma \tau _{QP}$ $=\eta ^{\prime }/\Delta \left(
T\right) $ is the probability of trapping a QP into a localized state and $%
n_{QP}(T)$ is the number of photoinduced QPs at temperature $T$ created by
each laser pulse. Since the number of photoexcited carriers is typically
small compared to the overall carrier density $\eta n_{QP}\ll n_{L}(0)$, we
can estimate $n_{L}(0)$ as

\begin{equation}
n_{L}(0)=\sqrt{\frac{\eta n_{QP}(T)}{Rt_{r}}}
\end{equation}

As a result, combining (4) and (7) we get an expression for the $T$%
-dependence of the photoinduced transmission (or reflection) amplitude {\em %
at low temperatures}:

\begin{equation}
\left| {\it \Delta }{\it T}/{\it T}\right| \propto n_{L}(0)=\sqrt{\frac{\eta
n_{QP}(T)}{\alpha t_{r}}}\frac{\Omega _{c}}{\Delta _{s}(T)}
\end{equation}

where $n_{QP}$ for a BCS-like case is given by \cite{kdbm}:

\begin{equation}
n_{QP}(T)=\frac{{\cal E}_{I}/({\bf \Delta }_{s}(T)+k_{B}T/2)}{1+\frac{2\nu }{%
N(0)\hbar \Omega _{c}}\sqrt{\frac{2k_{B}T}{\pi {\bf \Delta }(T)}}\exp (-{\bf %
\Delta }_{s}(T)/k_{B}T)}.
\end{equation}
Here ${\cal E}_{I}$ is the energy density deposited per pulse, $\nu $ is the
effective number of phonons per unit cell involved in the relaxation process
and $N(0)$ is the density of states at the Fermi energy in units eV$^{-1}$%
cell$^{-1}$spin$^{-1}$.

Near $T_{c},$ when the number of thermally excited localized carriers
becomes comparable or larger than number of nonequilibrium carriers, the
relaxation terms $2RN_{L0}n_{L}$ and $\tilde\gamma n_{L}$ become dominant,
and the solution to Eq. (3) has an exponential rather than a power law time
dependence of the form $n_{L}(t)=n_{L}(0)\exp \left( -t/\tau \right) ,$
where $1/\tau =\left( 2R(T)N_{L0}+\tilde{\gamma}\right).$ For $T\rightarrow
T_{c},$ the temperature dependence of the photoinduced signal amplitude is
then given by:

\begin{equation}
\left| {\it \Delta }{\it T}/{\it T}\right| \propto n_{L}(0)=\frac{\eta
n_{QP}(T)}{(2N_{L0}R(T)+\tilde{\gamma})t_{r}}
\end{equation}
Thus the predicted amplitude of the signal increases with increasing $T$ up
to $T_{c}$ and drops to zero above $T_{c}.$

In the case of a $T$-independent gap (or ''pseudogap''), $\Delta ^{p}$, we
assume that the gap exists at all temperatures. Since the gap does not close
at $T_{c}$, the recombination rate does not go to zero at $T_{c}$ and
instead of (4) we have a constant, $\Gamma $.

However, below $T_{c}$ the presence of the condensate may also have an
effect on the recombination of localized excitations. In general, the
relaxation rate is a function of the order parameter. To take this into
account, we can expand it in terms of even powers of $\Delta $. Near $T_{c}$
the order parameter is small and we can keep only the lowest power in $%
\Delta ^{2}$ \cite{Grisha}. If we assume that the order parameter exhibits
mean-field behavior $(\Delta \propto \sqrt{1-T/T_{c}}),$ then:

\begin{equation}
R\simeq \alpha (1-T/T_{c})+\Gamma
\end{equation}
where $\alpha $ is phenomenological constant which describes the dependence
of the relaxation rate on $T$ below $T_{c}$, which in general is not equal
to 0.

A slightly different expression for $R$ is obtained if we assume that the
recombination rate is dependent on the pair momentum in the condensate
through the kinetic energy. To illustrate this, we write the recombination
rate as $\Gamma =\Gamma _{0}+\bar{\Gamma}$ where $\bar{\Gamma}$ is the
momentum averaged recombination rate, and $\Gamma _{0}$ is the recombination
rate for pairs with $k=0$. The recombination rate is generally proportional
to the number of pairs in the condensate $n_{p}.$ For Bose condensation this
is given by $n_{p}\propto (1-(T/T_{c})^{3/2}),$ and we thus obtain a formula
for the total relaxation rate which is similar to Eq. (11), but with a
different temperature dependence in the first term: 
\begin{equation}
R=(\Gamma _{0}-\bar{\Gamma})(1-(T/T_{c})^{3/2})+\bar{\Gamma}
\end{equation}
In principle the two cases (11) or (12) can be distinguished by measurements
of the $T$-dependence of the photoinduced transmission or reflection below $%
T_{c},$ although the difference will be very small and very high quality
data is needed to do this.

To obtain the photoinduced signal amplitude, we substitute Eq.(11) into eq.
(3):

\begin{equation}
\left| {\it \Delta }{\it T}/{\it T}\right| \propto n_{L}(0)=\sqrt{\frac{\nu
n_{QP}^{\prime }}{(\alpha (1-(T/T_{c})^{\beta })+\Gamma )t_{r}}},
\end{equation}
where $\beta $ is a constant (generally 1 or 3/2) and $n_{QP}^{^{\prime }}$
is the equivalent of formula (9) for the case of a temperature independent
gap \cite{kdbm}:

\begin{equation}
n_{QP}^{^{\prime }}=\frac{{\it E}_{I}/{\bf \Delta }^{p}}{1+\frac{2\nu }{%
N(0)\hbar \Omega _{c}}\exp (-{\bf \Delta }^{p}/k_{B}T)}.
\end{equation}
In contrast to the BCS case, expression (13) is non-zero above $T_{c}$ and
reduces to: 
\begin{equation}
\left| {\it \Delta }{\it T}/{\it T}\right| \propto n_{L}(0)=\sqrt{\frac{\nu
n_{QP}^{^{\prime }}}{\Gamma t_{r}}.}
\end{equation}
which implies that the photoinduced absorption signal should remain
observable well above $T_{c}$ and should reveal the presence of a pseudogap
if it exists.

Just as before, a crossover to exponential time relaxation takes place when
the number of thermally excited excitations become large and the second term
in the Eq.(3) becomes dominant, i.e. $T\sim \Delta ^{p}$, leading to a
linear intensity dependence of $\left| {\it \Delta }{\it T}/{\it T}\right| $
given by modified Eq.(10) with $n_{QP}^{^{\prime }}$ and R from Eq.(12).
Note that this crossover may occur at higher temperatures than
experimentally measured since $T$-independent gap is typically $\Delta
^{p}\gg 300K.$

\section{Comparison with experimental data}

The temperature dependence of $\left| {\it \Delta }{\it T}/{\it T}\right| $
or $\left| \Delta {\cal R}/{\cal R}\right| $ for optimally doped YBa$_{2}$Cu$%
_{3}$O$_{7-\delta }$ \cite{Stevens,unpubdata} and for K$_{0.3}$MoO$_{3}$ 
\cite{bb} is plotted in Fig. 3a) and b) respectively. In both compounds the
signal amplitude increases with increasing temperature followed by an abrupt
drop above $T_{c}$. Now we compare the predicted $T$-dependences of $\left| 
{\it \Delta }{\it T}/{\it T}\right| $ (Eqs. (8) and (10)) with the data. The
parameters used in the fits are the same as previously used in the analysis
of the fast relaxation component \cite{kdbm,bb} with values of dimensionless
constant $\frac{2\nu }{N(0)\hbar \Omega _{c}}$ $\simeq 30$ for YBa$_{2}$Cu$%
_{3}$O$_{7-\delta }$ \cite{kdbm} and $\simeq 10$ for K$_{0.3}$MoO$_{3}$ \cite
{bb}. Since the magnitude of the gap is of the order of $\Delta _{c}\left(
0\right) $ $\sim 5kT_{c}$ we expect that Eq. (8) is valid up to temperatures
close to $T_{c}$. In Fig. 3a) and b) we show the calculated temperature
dependences of $\left| {\it \Delta }{\it T}/{\it T}\right| $ using Eq.(8) in
comparison with experimental data for YBa$_{2}$Cu$_{3}$O$_{7-\delta }$ and K$%
_{0.3}$MoO$_{3}$ respectively. We would like to stress that Eq.(8) is
independent of the shape of DOS of localized states within the gap and shows
a universal temperature dependence. It can be seen from the Fig.3b) that at
low temperature there is deviation of the calculated curve from the
experimental points. To explain this effect we should remember that Eq.(4)
for constant $R$ is valid near $T_{c}$ where $\Delta _{c}\left( T\right) $
is small. (If we add next - fourth order - term in the expansion of $R$ in
powers of $\Delta _{c}\left( T\right) $ we can account for this discrepancy.)

Near $T_{c}$ Eq.(8) fails and Eq. (10) should be used (see also Appendix).
It leads to a crossover from square root intensity dependence of $\left| 
{\it \Delta }{\it T}/{\it T}\right| $ at low $T$ described by Eq.(8) to
linear intensity dependence near $T_{c}$ predicted by Eq.(10).

Finally let us discuss the effect of $\tilde{\gamma}$. In Fig. 3a) and b)
fits to the data using the general solution (Eq.(19)) are shown with dashed
lines. In these fits we have also added the fourth order term in the
expansion of $R$ in powers of $\Delta _{c}\left( T\right) $. As can be seen
from these fits the effect of $\tilde{\gamma}$ becomes important near $T_{c}$
by cutting the divergence of $\left| {\it \Delta }{\it T}/{\it T}\right| $
as $T\rightarrow T_{c}$.

In Fig.3c) we have plotted calculated values of $\left| {\it \Delta }{\it T}/%
{\it T}\right| \propto n_{L}(0)$ for the case of temperature independent
pseudogap (which might be applicable in underdoped cuprates for example) as
a function of temperature for different values of parameter $\alpha $ using
Eq.(13). As can be seen from the Fig.3c) in this case slow relaxation via
localized states is present also above $T_{c}$. This effect is due to
temperature dependence of $n_{QP}^{^{\prime }}$ controlled by $T$%
-independent pseudogap above $T_{c}$.

In cuprate superconductors there is spectroscopic evidence suggesting that
there is a significant density of states in the gap possibly extending to
the Fermi level, which is often attributed to a $d$-wave gap symmetry.
However, by normal spectroscopies it is difficult to determine if the states
in the gap are QP states or, for example, localized states. Time-resolved
spectroscopy can answer this question rather effectively because of the
different time- and temperature-dependences of the QPs and localized carrier
relaxations. It was argued that in the presence of impurity scattering QP
DOS in the $d$-wave state remains finite at zero energy \cite{Gorkov}.
Recently it was proposed \cite{Lee} that the quasiparticles in the
superconducting state may become strongly localized for short coherence
length $d$-wave superconductors. However this statement has been questioned
by Balatsky and Salkola \cite{Balatsky} and remains controversial. On the
basis of available experimental data we cannot make any definite conclusion
about {\it origin} of intragap localized states.

We can, however estimate the density of the intra-gap states from the
available data by assuming that the optical probe process (step 3)\ is
similar for excited state absorption from localized states and for QPs. Both
optical probe processes involve allowed transitions to the same final state $%
E_{2}$ and so this assumption is not unreasonable. From typical photoinduced
reflection data for YBCO (as in Figure 1), we find that approximately $%
\left| {\it \Delta }{\it T}/{\it T}\right| _{L}\simeq \left| {\it \Delta }%
{\it T}/{\it T}\right| _{QP}$, implying that also $n_{L}(0)\simeq n_{QP}.$
From this we can conclude that the density of intra-gap states is {\em %
comparable} with the density of QP states. This observation has important
implications for the interpretation of frequency-domain spectroscopies,
since it suggests that the spectra should show a very significant intra-gap
spectral density due to localized states, irrespective of the gap symmetry.

Assuming that the optical transition probability of the probe pulse is the
same for QPs as for the intra-gap states, from Eqs. (8) and (9) we obtain: 
\begin{equation}
\frac{\left| {\it \Delta }{\it T}/{\it T}\right| _{L}}{\left| {\it \Delta }%
{\it T}/{\it T}\right| _{qp}}\sim \frac{n_{L}}{n_{QP}}=\frac{\eta \tau _{L}}{%
t_{r}}
\end{equation}
where $\left| {\it \Delta }{\it T}/{\it T}\right| _{qp}$ is the photoinduced
transmission due to the QPs. From Fig. (1) typically $n_{L}/n_{QP}=0.1-1$
and using a pulse repetition rate $t_{r}$ = 12 ns and with $\tau _{L}=100$
ns, we obtain an estimate of the trapping probability for carriers by
localized states of $\eta =0.1-1$.

A detailed discussion of the origin of the localized intra-gap states in the
cuprates should be deferred until more systematic data as a function of
doping is available, and we only mention some of the most likely
possibilities: (i) localized states associated with the inhomogeneous ground
state of the cuprates (stripes) \cite{stripes}, (ii) intrinsic defect
states, (iii) localized QP states in $d$-wave superconductor \cite{Lee} and
possibly (iv) holons \cite{Anderson}. In K$_{0.3}$MoO$_{3}$, the nature
intra-gap excitations has been a subject of extensive study over the years
and the reader is referred to ref. \cite{Gruener} for a review. However, the
fact that the signals in K$_{0.3}$MoO$_{3}$ and YBa$_{2}$Cu$_{3}$O$%
_{7-\delta }$ are very similar appears to rule out both spin excitations and
vortex states, leaving localized charges as the most plausible origin of the
intra-gap states.

\section{Conclusions}

To conclude, the calculated time- and temperature- dependence of the
photoinduced absorption for the case of a BCS-like gap is found to be in
good agreement with experimental data from femtosecond time-resolved
spectroscopy on the cuprate superconductor YBa$_{2}$Cu$_{3}$O$_{7-\delta }$
near optimum doping and the charge-density wave insulator K$_{0.3}$MoO$_{3}$%
. We find that time-resolved spectroscopy can very effectively distinguish
between QP states and localized states in the gap. A rather surprising
feature of the data is the remarkable separation of the QP response on the
femtosecond timescale and the slow response of intra-gap state relaxation on
the scale of 100s of nanoseconds. In both materials we find a significant
intra-gap density of states, which display very different time-dynamics and $%
T$-dependence than the QP states above the gap.

The authors wish to acknowledge the ULTRAFAST network and the Ministry for
Science and Technology of Slovenia for supporting part of this work.

\section{Appendix}

Analytic solution of Eq. (3) has the form: 
\begin{equation}
n_{L}(t)=\frac{2N_{L0}C\exp (-t/\tau )}{1-C\exp (-t/\tau )}.
\end{equation}
Here $1/\tau =2N_{L0}R+\tilde{\gamma}$. Constant $C$ can be found from the
following equation (see also Eq.(6)): 
\begin{equation}
n_{L}(0)-n_{L}(t_{r})=\eta n_{QP}.
\end{equation}
Combining this two equations one obtains the following form for $n_{L}(0)$: 
\begin{eqnarray}
n_{L}(0) &=&N_{L0}(1+\tilde{\gamma}/2RN_{L0}) 
\nonumber  %
%
\\
&&\left[ \sqrt{1+\frac{\eta n_{QP}}{N_{L0}^{2}(1+\tilde{\gamma}%
/2RN_{L0})^{2}Rt_{r}}}-1\right]
\end{eqnarray}
This solution reduces to Eq.(8) if $\frac{\eta n_{QP}}{N_{L0}^{2}(1+\tilde{%
\gamma}/2RN_{L0})^{2}Rt_{r}}\gg 1$ and to Eq.(10) in the opposite limit.

\medskip \bigskip 
\vrule height.2mm width7cm depth0.1pt%
%
\medskip \bigskip

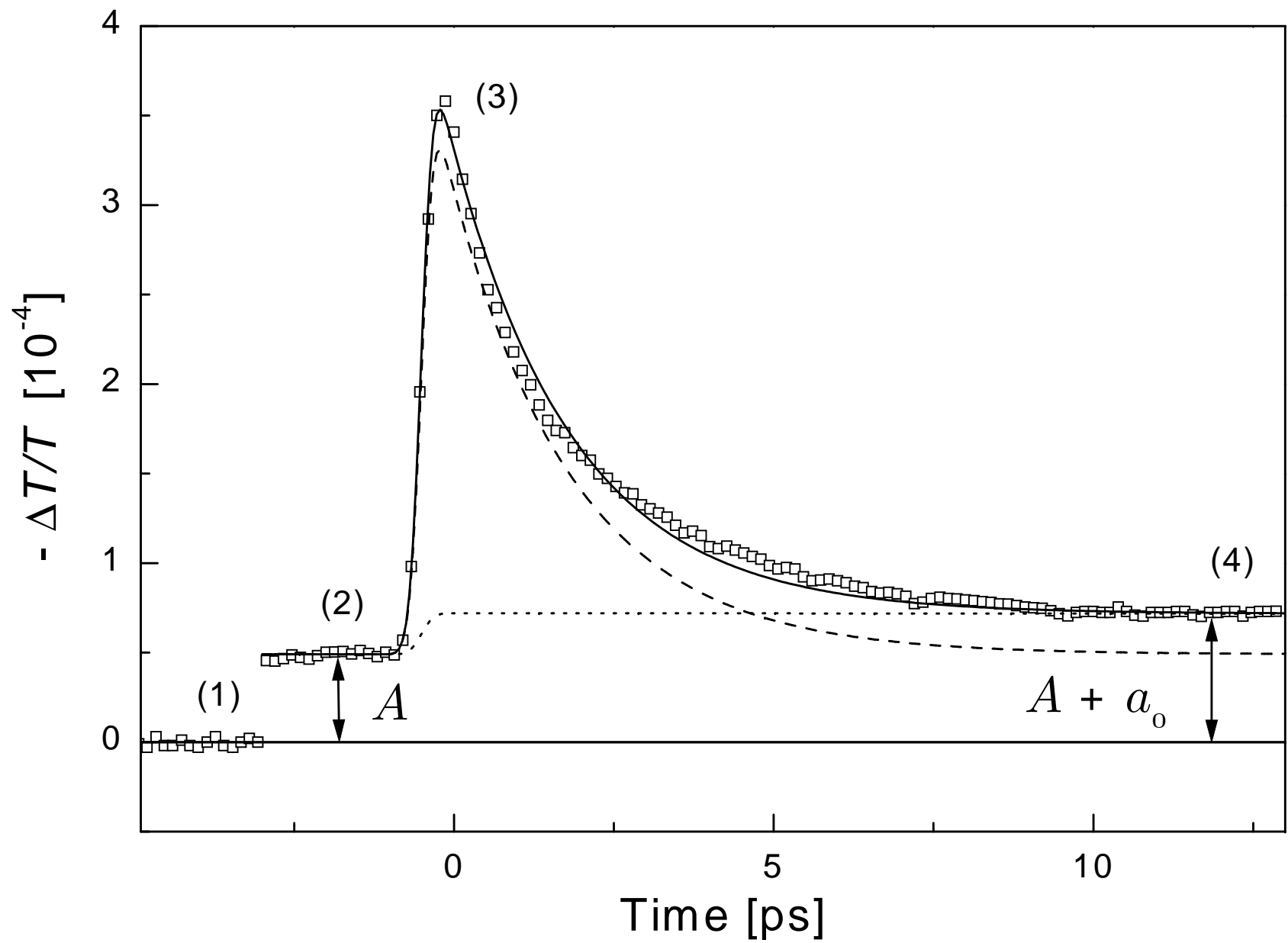
Figure 1. A photoinduced transmission signal ${\it \Delta }{\it T}/{\it T}$
as a function of time $t$ after photoexcitation in YBa$_{2}$Cu$_{3}$O$%
_{7-\delta }$ ($T_{c}=90$K) taken at $T$ = 80 K (points), together with the
fit (solid line). (1) is the baseline signal with no pump applied. (2) is
the long-lived signal pile-up remaining from previous pulses, (3) (dashed
line) is the signal due to QP recombination and (4) (dotted line) is the
long-lived signal remaining after all the QP signal has decayed. Signal
pile-up contribution and the single pulse contribution to the slow
photoinduced signal are given by $A$ and $a_{0}$ respectively.

Figure 2. a) The pump-probe optical diagram. {\sf 1} represents the pump
pulse exciting\ charge carriers into a higher-lying band, {\sf 2} the
carriers rapidly relax their energy to states near the Fermi energy. {\sf 3}
represents the probe pulse. b) A schematic diagram of the terms contributing
to the relaxation of intra-gap states in a superconductor with a gap 2$%
\Delta .$ In the case of a CDW gap, the $e-h$ pairs take the place of Cooper
pairs, so the rate equation remains the same.

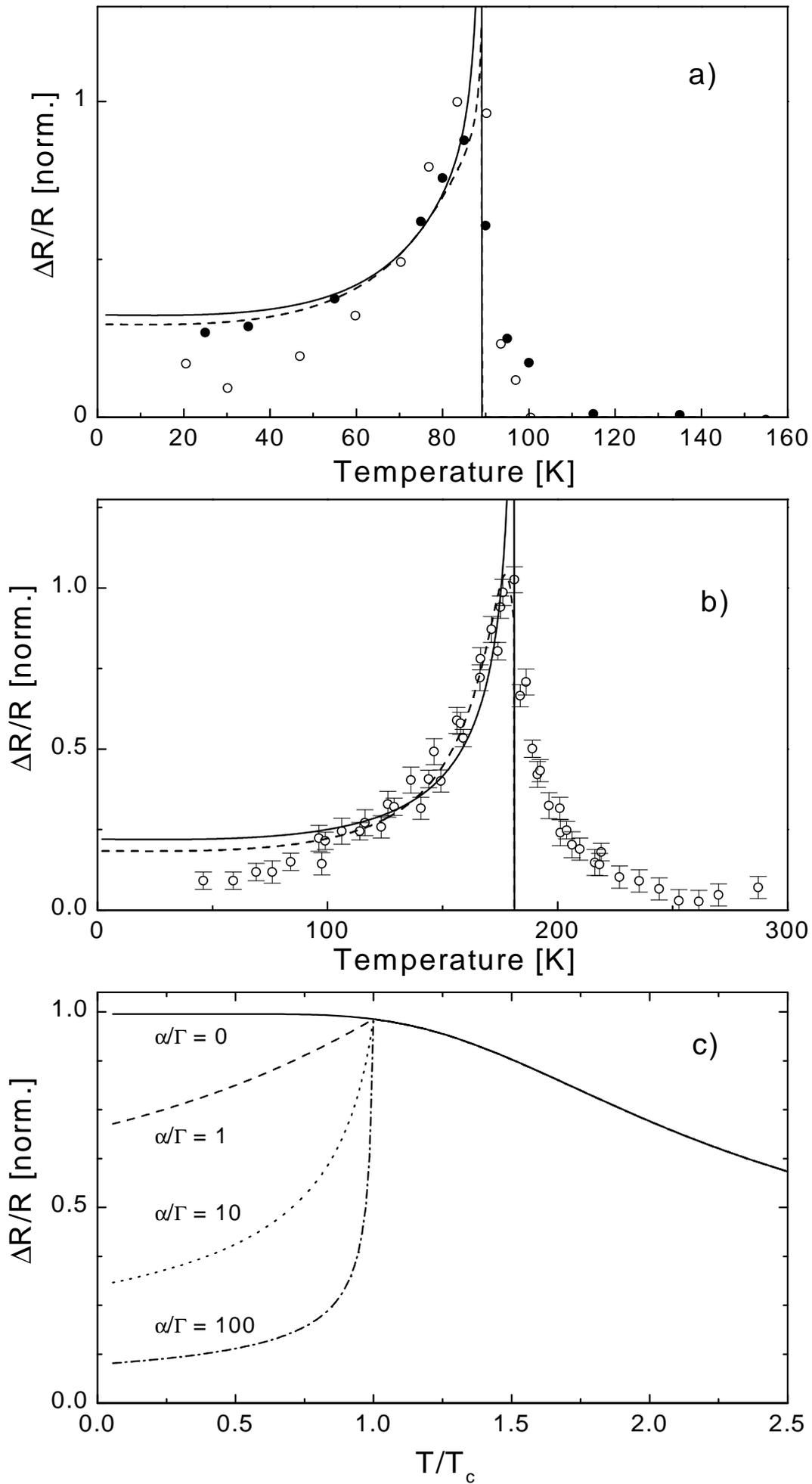
Figure 3. a) The temperature dependence of the photoinduced absorption from
localized states in YBa$_{2}$Cu$_{3}$O$_{6.9}$ taken from Ref. \cite{Stevens}
(open circles) and Ref. \cite{unpubdata} (solid circles). The solid line is
a plot of expression (8) with $\Delta _{c}\left( 0\right) /k_{B}T_{c}=5$
whereas the dashed line represents a general solution (Eq. 19) with non-zero 
$\tilde{\gamma}$ term and additional fourth order term in the expansion $R$
in powers of $\Delta _{c}\left( T\right) .$ b) The temperature dependence of
the photoinduced reflection in K$_{0.3}$MoO$_{3}$ from Ref.\cite{bb} (open
circles) compared with the model fit using Eq.(8) with $\Delta _{c}\left(
0\right) /k_{B}T_{c}=4.8$ (solid line). The general solution (Eq.(19)) is
represented by the dashed line. c) The calculated temperature dependence of
the photoinduced absorption from localized states in case of \ T-independent
pseudogap Eq.(13) with $\Delta ^{p}/k_{B}T_{c}=8$ and different $\alpha
/\Gamma $ ratios.

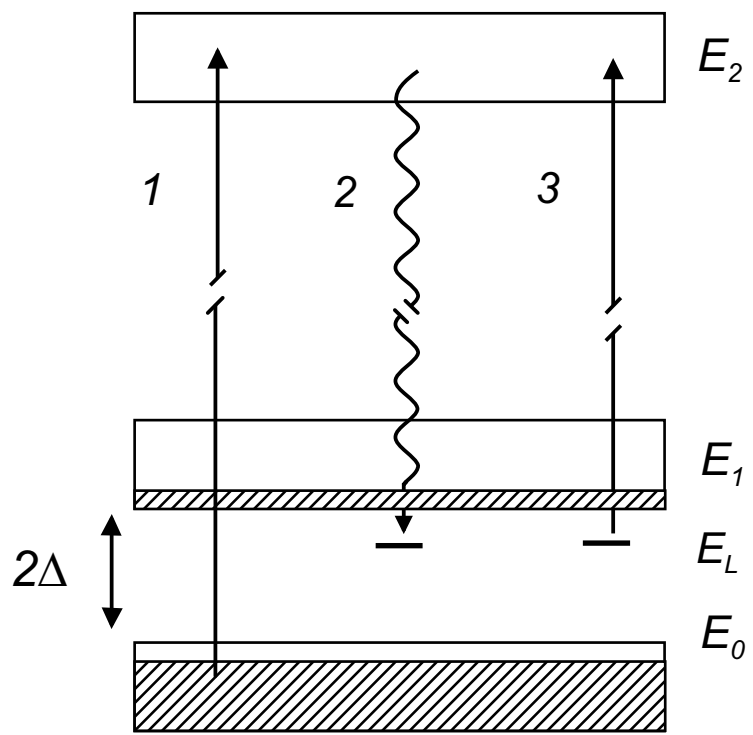
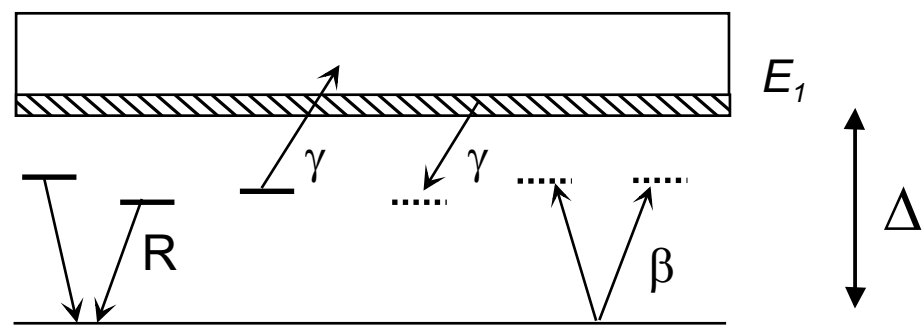
\end{document}